\documentclass{article}

\begin{document}

\author{C. S. Unnikrishnan$^{1,2,a}$, G. T. Gillies$^{3,b}$ \\
$^{1}$\textit{Gravitation Group, Tata Institute of Fundamental Research, }\\
\textit{Homi Bhabha Road, Mumbai - 400 005, India}\\
$^{2}$\textit{NAPP Group, Indian Institute of Astrophysics,}\\
\textit{Koramangala, Bangalore - 560~034, India}\\
$^{3}$\textit{School of Engineering and Applied Science, }\\
\textit{University of Virginia}, \textit{Charlottesville, VA 22904-4746}}
\title{Nano-constraints on the spatial anisotropy of the Gravitational
Constant}
\date{}
\maketitle

\begin{abstract}
We present constraints from various experimental data that limit any
spatial anisotropy of the Gravitational constant to less than a part per
billion or even smaller. This rules out with a wide margin the recently
reported claim of a spatial anisotropy of $G$ with a diurnal temporal
signature.
\end{abstract}

\medskip



\bigskip\ 

The question whether the gravitational constant has some spatial anisotropy
has been discussed in the context of nonstandard theories of gravity as well
as in the context of experimental observations\cite{will,goodkind}. The
general setting for discussing this issue in the Parametrized Post-Newtonian
framework is discussed extensively by Will\cite{will}. There are severe
constraints on the anisotropy parameters from gravimetric experiments\cite%
{goodkind} and Lunar Laser Ranging\cite{llr}. These anisotropies depend on
the magnitude and direction of the velocity of the frame in which the
experiments are done, with respect to a preferred frame. In the usual frame
work of alternate theories of gravity, a spatial anisotropy of $G$ that
depends merely on the relative angle between the line of interaction of two
gravitating bodies and a reference direction with respect to the distant
stars is not expected. However, recently there has been a claim for the
observation of a significant anisotropy in $G$ from an experiment that
monitored the gravitational interaction between a source mass and the mass
element of a torsion balance over a period of 7 months \cite{gersht}. The
reported result was that $G$ changes with orientation of the line of
interaction of the masses by at least 0.054\%. This is a very large
anisotropy, and of a different kind, when contrasted with what is usually
discussed in the context of theoretical models. In this paper we discuss
possible constraints on such an anisotropy and show that results from
several standard experiments that are sensitive to the gravitational
interaction between the earth and another body constrain any spatial
anisotropy of $G$ to less than a part in a billion. This rules out the
reported claim of the spatial anisotropy of 0.054\% by a factor of more than
a million. While our discussion focuses on a recent experiment as an example%
\cite{gersht}, the constraints we present are very general and apply in a
model independent way to the issue of spatial anisotropy of $G$ that depend
on the direction of the line of interaction with respect to a preferred
frame.

Theoretically, the issue of spatial anisotropy of $G$ is related to how
nearby masses and their distribution can affect the gravitational
interaction between two bodies, and also to preferred frame effects. If the
modification of the gravitational interaction depends on the gravitational
potential generated by other masses, then the most distant masses dominate
and the spatial anisotropy is expected to be small. The galactic
contribution to the anisotropy is then the dominant term in any reasonable
expectation, and the order of magnitude is then of the order of $%
\delta\approx GM_{g}/c^{2}R_{g}\approx10^{-6}.$ Anisotropy of even this
level can be expected only in very special theoretical scenarios since $%
\delta$ is the value of the scalar potential and it does not single out a
direction in space. Theoretically, existence of a preferred frame would give
rise to the dependence of $G$ on the velocity with respect to this frame and
such effects have been discussed extensively and completely in the context
of the PPN formalism \cite{will}. It is appropriate to summarize the
relevant physical issues in the context of the PPN formalism. Local physics
in General Relativity itself does not contain any dependence on the
location, orientation and the velocity of a frame, except the standard
couplings to gradients in the gravitational field (tidal forces). In
particular, $G$ is truly a constant in space and time. This is the case in
any theory that contains only the metric to represent long range fields. In
a metric theory that contains a scalar field in addition (Scalar-Tensor
theory), $G$ can depend on location and time through the variations in the
scalar field, but not on velocity and orientation of the frame. In a more
general type of theory, though, there could be local effects that depend on
position, velocity and orientation and even perhaps angular velocity (spin).
Note that any such effects point to a violation of the Strong Equivalence
Principle, and it is in this context we analyze the possible variation of $G$
with orientation. \ 

Experimentally the most common approach has been to look for such
anisotropies as the laboratory reference frame assumes different velocities
and positions as the earth moves with respect to a nonrotating reference
frame fixed to the galactic centre. If one is searching for an effect that
depends on the orientation of the line of interaction between two bodies,
then the signal is typically of the diurnal period and its second harmonic.
The line of interaction sweeps through a fixed plane containing the rotation
axis of the earth twice during a diurnal period. \ Thus the primary signal
should be at twice the diurnal frequency (semi-diurnal period), like a tidal
force, and a signal at the diurnal period could also be present due to
possible asymmetry in the second harmonic signal because of the inclination
of the earth's rotation axis, for example, with respect to the galactic
plane. This requires that the experimental conditions are controlled and
monitored with great care since there are many sources -- variations in
ambient conditions like temperature, pressure and magnetic field,
electromagnetic couplings, human interferences etc. -- that can generate a
large diurnal signal and its second harmonic in high sensitivity
experiments. It is important to note that detecting variations in $G$ with a
relative precision better than $10^{-5}$ is beyond the current experiments
that measure the absolute value of $G$\cite{rop}. But, measuring variations
in $G$ relative to fixed average value can be done with a precision better
than $10^{-9}$, and this is what allows us to constrain the spatial
anisotropy of $G$ to better than 1 part in a billion. A measurement of $%
\Delta g/g$ is equivalent to a measurement of $\Delta G/G$ in such
experiments, after accounting for known effects that cause a nonzero $\Delta
g/g.$ This point has been noted long ago for constraining PPN parameters\cite%
{will}.

The Hughes-Drever type very high precision null experiments test the
anisotropy of inertia\cite{hughes}. They however do not test a spatial
anisotropy of $G$ since the anisotropy of inertia affects the spectral line
splitting through a nongravitational interaction involving the ratio of
charge to inertial mass.

The experiment by Gershteyn \textit{et al} consisted of repetitive
measurements of $G$, separated by a time interval of about 1.2 hours. \ They
measured the change in the period of a torsion pendulum in the presence of a
heavy source mass\cite{gersht}. The measurements were repeated round the
clock, for a period of about 7 months. The periodogram showed a strong peak
at the diurnal period and this was interpreted as evidence for a spatial
anisotropy of the gravitational constant amounting to larger than $5\times
10^{-4}$, revealed as the line of interaction of the masses swept around
with the rotating earth with respect to the distant stars.

This is easily shown to be not due to a spatial anisotropy of the
gravitational constant. There are several reasons that immediately reject
the hypothesis that the observed signal was due to an anisotropy in $G$
detected as the line of interaction of the masses assumed different
orientations with respect to a frame defined by distant stars. There was not
a strong and significant signal at the semi-diurnal period (some of their
data contains significant power at the semi-diurnal period, but the noise in
the nearby bins are also large). Though this in itself could be considered
as evidence that the signal in their experiment is spurious and not an
indication of $G$-anisotropy, strong numerical constraints that we discuss
below conclusively prove that \emph{any} $G$-anisotropy signals above $%
\Delta G/G\simeq10^{-10}$ should be considered spurious.

Before we discuss a stringent limit, we illustrate how to easily obtain a
limit good enough to rule out anisotropy of $G$ at the level of $10^{-6},$
and thus conclusively rule out the result claimed by Gershteyn et al. A
simple experiment that can be performed in almost any laboratory using a
commercial electronic weighing balance is capable of constraining anisotropy
of $G$ to below $10^{-6}.$ Commercial electronic balances (like the Mettler
AE series) have sensitivity of about 10 to 100 microgram for measuring
masses in the range of 10 to 500 grams. Even after allowing for long term
drifts (whose diurnal component will mimic a signal), these balances can
measure weight changes with a precision better than $10^{-6}$, especially if
the signal is repetitive with a specific period, like the diurnal period.
Since the weight measured is directly proportional to the local
gravitational attraction of the earth on the sample being weighed, the
fractional change in weight is equal to the fractional change in the
gravitational constant. At the sensitivity of $10^{-6},$ tidal effects from
the solid deformation of the earth or ocean loading are not observed, and
there is no need for correction from tidal terms. Clearly, such an
experiment looks directly for a spatial anisotropy of the gravitational
constant with a sensitivity of $10^{-6}$ as the line joining the sample mass
and earth's centre sweeps out different directions with respect to the
galactic centre. We monitored a weight of about 150 grams for a few days and
did not observe any periodic component at the sensitivity limit of $100$ $%
\mu $g. This means that this simple experiment gives the remarkable
constraint $\Delta g/g\leq 6\times 10^{-7},$ which translates to a similar
constraint on $\Delta G/G$ with respect to orientation changes in the
galactic plane, since the experiment is done at a relatively small latitude
of \ 20 degrees. Absence of a signal at the diurnal period and its second
harmonic rules out the possibility that what Gershteyn et al observed was a
spatial anisotropy of $G$ by a wide margin -- a factor of about 1000.

The simple yet significant constraint of $\Delta G/G<10^{-6}$ we pointed out
can be improved by a factor of 1000 or more using other standard
experimental data, especially the kind obtained routinely from
microgravimeters with a sensitivity exceeding 1 microgal ($10^{-8}$ m/s$%
^{2}\simeq 10^{-9}$ g). All one needs is the residual diurnal and
semi-diurnal component in the local gravitational field value after
subtracting out the known tidal contributions. We had already used such data
to constrain the Majorana shielding factor earlier \cite{ug,ug2}. We had
analyzed 2 weeks' continuous gravimetric data from a Chinese experiment \cite%
{chinese}, obtained bracketing a solar eclipse, and found that the diurnal
signal is smaller than 0.06 microgal. The signal at the second harmonic is
even smaller. This translates to a fractional diurnal change in local
gravity, after subtracting known tidal effects, smaller than $\frac{\Delta g%
}{g}\leq 6\times 10^{-11}.$ The experimental site was at a latitude of about
53 degrees, and the angle factors reduce the sensitivity of the experiment.
(This would be the case if significant contributions are assumed from
orientations in the galactic plane). Even allowing for a large sensitivity
reduction of 0.1, we can conservatively obtain the constraint that $\Delta
G/G<6\times 10^{-10}.$ This constraint is almost a million times smaller
than the spatial anisotropy claimed by Gershteyn et al, and therefore there
is no doubt that they are observing a spurious systematic effect.

This constraint can be improved by another factor of 10 without much effort
using data from a long term microgravimetry experiment at suitable
locations. In fact, there is already data from a dedicated and pioneering
experiment done long ago, using a superconducting gravimeter, that rules out
such an anisotropy at the level of \ $\Delta G/G<10^{-10}.$ This experiment
by Warburton and Goodkind\cite{goodkind} collected gravimetric data over an
18 month duration with sub-microGal sensitivity and put constraints on the
PPN anisotropy parameters at the level of 0.1\%. A detailed analysis of the
data at various frequencies and phases of relevance was done. The amplitude
of the residuals after subtracting the known tidal terms at diurnal and
semi-diurnal periods was at most 0.3 microGal and 0.1 microGal respectively,
and phase analysis constrains this even better. Thus a model independent
constraint at the level of $\Delta G/G\leq10^{-10}$ can be obtained from
their data. In specific models with the galactic mass as the source of the
anisotropy, phase analysis allows tighter constraints of $\Delta
G/G\leq10^{-11}$ (the residual amplitude used for constraining the PPN
preferred frame parameter corresponded to about 0.006 microGal in their
data).

There is another source of data that constrains spatial anisotropy of $G$
even better. The Lunar Laser Ranging (LLR) data has been used for
constraining the fractional temporal variation of the gravitational constant
to about $4\times10^{-12}$/year\cite{llr}$.$ The same analysis then also
constrains the spatial anisotropy to similar levels, since a periodic signal
with a period close to the lunar month (corrected for the earth yearly
motion) will show up if there is spatial anisotropy in $G$, as the line
joining earth and moon sweeps out different directions with respect to the
direction to the galactic centre. The sensitivity with which a periodic
signal can be pulled out is of the same order (or sometimes better) than the
sensitivity with which a long term drift can be analyzed. Absence of such a
signal then constrains $\Delta G/G<4\times10^{-12}$ or so in the galactic
neighbourhood. This remarkable constraint cannot be easily surpassed by
laboratory experiments.

A large variety of other experiments could also be cited as evidence against
the anisotropy claimed by Gershteyn et al. See the reviews of Gillies\cite%
{rop} and Unnikrishnan and Gillies\cite{ug2} for some further discussion of
anomalous gravitational effects and see the bibliographic listing complied
by Gillies\cite{gillies} for citations to relevant experiments. While it is
difficult to say what the origin of the effect might be in the results of
Gershteyn et al., it is possible to consider some interesting alternatives.
Their experimental apparatus has a long history of use for the measurement
of the gravitational constant, see eg., \cite{karaold,karanew}. Their
torsion pendulum system has been used for that purpose in both symmetric and
asymmetric modes of operation, i.e., with the small mass system in the
gravitational potential of either two attracting masses or just one,
respectively. For the data taken in the experiments under consideration
here, they used their system in the asymmetric mode, such that the dumbbell,
which is a mass quadrupole, was coupled gravitationally to the point-source
field of the large spherical attracting mass, which is a single mass
monopole. Variations in the ambient environment as well as changing
gravitational gradient couplings could be responsible for a signal at the
diurnal period. We note that the most recent high-precision measurements of $%
G$ incorporate mass distributions such that the gravitational interaction is
as well-defined as possible in order to avoid such problems; see eg. \cite%
{gund}. However, absolute $G$-measuring experiments have not yet achieved
the sensitivity to probe a variation in G less than $0.001\%$, though
several experiments have recently reached enough sensitivity to probe the
effect claimed by Gershteyn et al with a signal to noise ratio of about 10%
\cite{gund,quinn}.

We conclude by stressing that any claim for a spatial anisotropy of the
gravitational constant that depends on the direction of line of interaction
with respect to a fixed frame should be considered as spurious if reported
to be larger than about $10^{-10},$ which is a conservative constraint
obtained from laboratory experiments.

\medskip\ 

\noindent \textbf{Acknowledgements}: We thank R. Cowisk for suggestions on
possible new experiments to probe anisotropy of $G$. The experiment using
the Mettler balances were done with P. L. Paulose at TIFR, Bombay. C. S. U
thanks the Department of Science \& Technology, Government of India for
financial support for a Non-accelerator particle physics program. G. T. G.
received partial financial support from a NASA subcontract from the
University of Tennessee to the University of Virginia.

\bigskip\ \ \ \ 

\noindent E-mail addresses: $^{a}$unni@tifr.res.in, $^{b}$gtg@virginia.edu\

\end{document}